\documentclass{aa}  

\usepackage{graphicx}
\usepackage{xcolor}
\usepackage{txfonts}
\usepackage{booktabs}
\begin{document} 
 
\title{{\it Hubble} Space Telescope survey of Magellanic Cloud star clusters. 
 Binaries, mass functions, blue stragglers, and structural parameters}

   \author{F. Muratore \inst{1}, 
          M. V. Legnardi \inst{1}, 
        A. P. Milone \inst{1,2}, 
        G. Cordoni \inst{3}, 
        A. Mastrobuono-Battisti \inst{1,2,4},
        A. F. Marino \inst{2}, 
        T. Ziliotto \inst{5},
        E. Dondoglio \inst{6}, 
        E. Bortolan \inst{1}, 
        E. P. Lagioia \inst{7} 
        }

   \institute{Dipartimento di Fisica e Astronomia “Galileo Galilei”, Università Degli Studi di Padova, Vicolo dell’Osservatorio 3, 35122 Padova, Italia
     \and
     Istituto Nazionale di Astrofisica - Osservatorio Astronomico di Padova, Vicolo dell’Osservatorio 5, 35122 Padova, Italy 
     \and 
     Research School of Astronomy and Astrophysics, Australian National University, Canberra, ACT 2611, Australia  
      \and
    Dipartimento di Tecnica e Gestione dei Sistemi Industriali, Università degli Studi di Padova, Stradella S. Nicola 3, I-36100 Vicenza, Italy  
        \and
        NSF’s NOIRLab, 950 North Cherry Avenue, Tucson, AZ 85719, USA 
        \and
        Physics Department, American University of Sharjah, P.O. Box 26666, Sharjah, UAE
        \and
    South-Western Institute for Astronomy Research, Yunnan University, Kunming 650500, PR China 
      \\    
}

   \date{Received May XX, XXXX; accepted May XX, XXXX}
\titlerunning{Binaries, mass functions and structural parameteres} 
\authorrunning{Muratore et al.}

\abstract
{Binary stars are key tracers of the dynamical evolution of star clusters and provide important constraints on stellar populations and mass functions. The Magellanic Clouds host clusters spanning a wide range of ages and masses, offering an ideal laboratory to investigate these properties in regimes poorly sampled in the Milky Way.
We aim to characterize the binary populations, mass functions (MFs), blue straggler (BS) content, and structural parameters of intermediate-age Magellanic Cloud clusters, and to explore their dependence on global cluster properties.
We analyze high-precision \textit{Hubble} Space Telescope photometry for 16 clusters obtained with ACS/WFC and WFC3/UVIS. 
Structural parameters are derived from stellar density profiles. Binary fractions are measured using the \textit{binary map} technique, focusing on systems with mass ratios $q>0.7$. We derive MFs accounting for unresolved binaries and identify candidate BS populations from color-magnitude diagrams.
The fraction of binaries with $q>0.7$ ranges from 5\% in NGC \,2121 up to 13\% in NGC \,2173, with a mass-ratio distribution that is consistent with being flat on average. By combining our results with literature data, we confirm a clear anti-correlation between the core binary fraction and cluster mass, while no significant dependence on cluster age is found. 
The clusters follow the established relation between age and core radius, although with substantial scatter at fixed age. Within the narrow age range explored here, clusters exhibiting steeper MFs are found to have smaller core radii. We find no evidence for a correlation between the fractions of binaries and BS fractions.
These findings are consistent with a scenario in which dynamical evolution plays a primary role in the formation of binary populations. The connection between MF slope and structural parameters provides new constraints on cluster evolution and suggests a link between MF slope and structural evolution.}


   \keywords{Young stellar cluster -- Stellar populations -- Magellanic clouds -- Binary stars -- Mass functions}

   \maketitle

%

\section{Introduction}

Binary stars play a fundamental role in the evolution and formation of star clusters. 
For instance, binary populations provide valuable constraints on the dynamical evolution of stellar systems, since they act as a significant source of energy. Additionally, determining the binary fraction is crucial for interpreting cluster dynamics, characterizing stellar populations, and constraining possible cluster formation pathways.

In high-density stellar environments, frequent close encounters promote the disruption of binary systems, whereas in low-density environments the binary fraction is expected to increase \citep{kaczmarek2011, deacon2020}. The sensitivity of binary star systems to environmental conditions makes them valuable for investigating the multiple stellar population phenomenon in globular clusters (GCs). These ancient systems are composed mainly of a first and a second stellar population that differ in their chemical compositions. 
The prevalence of binaries in these subpopulations provides important information on the formation channel of multiple stellar populations \citep[e.g.][]{vesperini2011,hong2015,hong2016,sollima2022,Hypki2022,bortolan2025}.

Understanding binary fractions in stellar systems has implications beyond cluster dynamics. 
The high frequency of stellar encounters makes star clusters ideal environments for the formation of exotic objects. The study of exotic stellar objects, such as blue stragglers (BSs), cataclysmic variables, millisecond pulsars, and low-mass X-ray binaries, has revealed their presence in close binary systems. This suggests that these objects are formed primarily through interactions within binary stars \citep{leigh2013,rain2024}.

Binary stars are also critical for interpreting several complex observational features of young star clusters. 
For instance, almost all clusters younger than 600 Myr show extended main sequence turn-offs and split main sequences, features that are consistent with stellar populations having different rotation rates \citep{milone2015a,dantona2015, milone2018,cordoni2018, marino2018a}. One suggested formation channel involves binary systems that can initially brake rapidly rotating stars, and \cite{muratore2024} demonstrated this crucial role in reproducing the observed properties of young clusters in Magellanic Clouds (MC).
Additionally, robust determinations of mass functions (MFs) in stellar systems require accurate constraints on the binary fraction \citep[e.g.,][]{legnardi2025}, since unresolved binaries can introduce substantial biases in the derivation of this relation.

Although extensive work has been carried out on binary populations, MFs, and other properties of Galactic clusters \cite[e.g.,][]{milone2012, milone2016, cordoni2023}, several key questions remain open. In particular, it remains unclear whether the binary fraction and, more broadly, cluster properties such as the MF and structural parameters, systematically depend on fundamental quantities such as the age and mass of the host cluster.

The seminal work of \cite{milone2012} presented the most comprehensive photometric analysis of binaries in 59 Galactic GCs using \textit{Hubble} Space Telescope (HST) Advanced Camera for Surveys (ACS) data. Their study revealed that binary fractions in GCs are typically lower than those in the Galactic field \citep[e.g.][]{duquennoy1991, fischer1992}, ranging from approximately 3\% to 38\%. 
However, the Galactic GC sample is relatively homogeneous in age, with most clusters older than ~10 Gyr, limiting our ability to study how binary properties vary with the age of the cluster and the formation environment.

Similarly, \cite{cordoni2023} presented a large survey of 78 Galactic open clusters (OCs) to investigate their binary populations, MFs, and BS populations. They reported overall binary fractions ranging from about $\sim$15\% up to more than $\sim$60\%, and MFs exhibiting comparable behavior, which can be accurately described by two single power-law functions, with a change in slope occurring around 1 $M_{\odot}$. Although their study includes clusters spanning a broad range of ages and masses, those in the interval 1–3 Gyr and $10^4$–$10^5$ $M_{\odot}$ remain insufficiently examined.

In this context, the MC provide a crucial laboratory to overcome these limitations, as they host stellar clusters spanning a wider and more continuous range of ages and masses, particularly in the parameter space that is poorly sampled in the Milky Way. \cite{mohandasan2024} presented a systematic photometric study of binaries in fourteen Magellanic Cloud star clusters with ages between ~0.6 and 2.1 Gyr and masses of $10^4-10^5 M_{\odot}$. Using HST ACS and Wide Field Camera 3 (WFC3) observations, they measured binary fractions for systems with mass ratios greater than 0.7, finding values ranging from ~7\% in NGC 1846 to ~20\% in NGC 2108 in the core region. When combined with results from OCs and GCs, their analysis reveals an anti-correlation between core binary fraction and cluster mass for masses greater than $10^5 M_{\odot}$, while clusters of mass smaller than $10^4 M_{\odot}$ exhibit a wide range of binary fractions, suggesting significant variations in initial binary content or subsequent dynamical evolution. However, this trend presents a lack of clusters covering the masses from $10^4$ to $10^5 M_{\odot}$, partially populated by the sample of \cite{mohandasan2024}, but still not enough to confirm the presence of this trend in MC.

In this work, we seek to explore clusters that cover a wide range of ages (1.2--2.9 Gyr) and masses ($10^4-10^5\,M_{\odot}$), and are located in environments that have been poorly studied. We analyzed the structural parameters of 16 star clusters in the Large and Small Magellanic Clouds (LMC and SMC, respectively), which are crucial for theoretical modeling. We extended the methodology introduced by \citet{muratore2026} to examine their binary populations. Additionally, we explored their MFs for the first time in this mass range for Magellanic Cloud clusters, as well as their BS populations.

This paper is organized as follows: Sect. 2 presents HST observations and photometric analysis techniques. It also provides details on the artificial star tests that were performed and on the procedure adopted to derive the density profile. Sect. 3 presents the methodology used to measure the binary fraction, derive the BSs fraction, and determine the MFs. Sect. 4 reports our results and comparisons with other surveys.

\section{Catalogs and Artificial stars}\label{sec:data}
\begin{figure*}
    \centering
    \includegraphics[width=1.\linewidth]{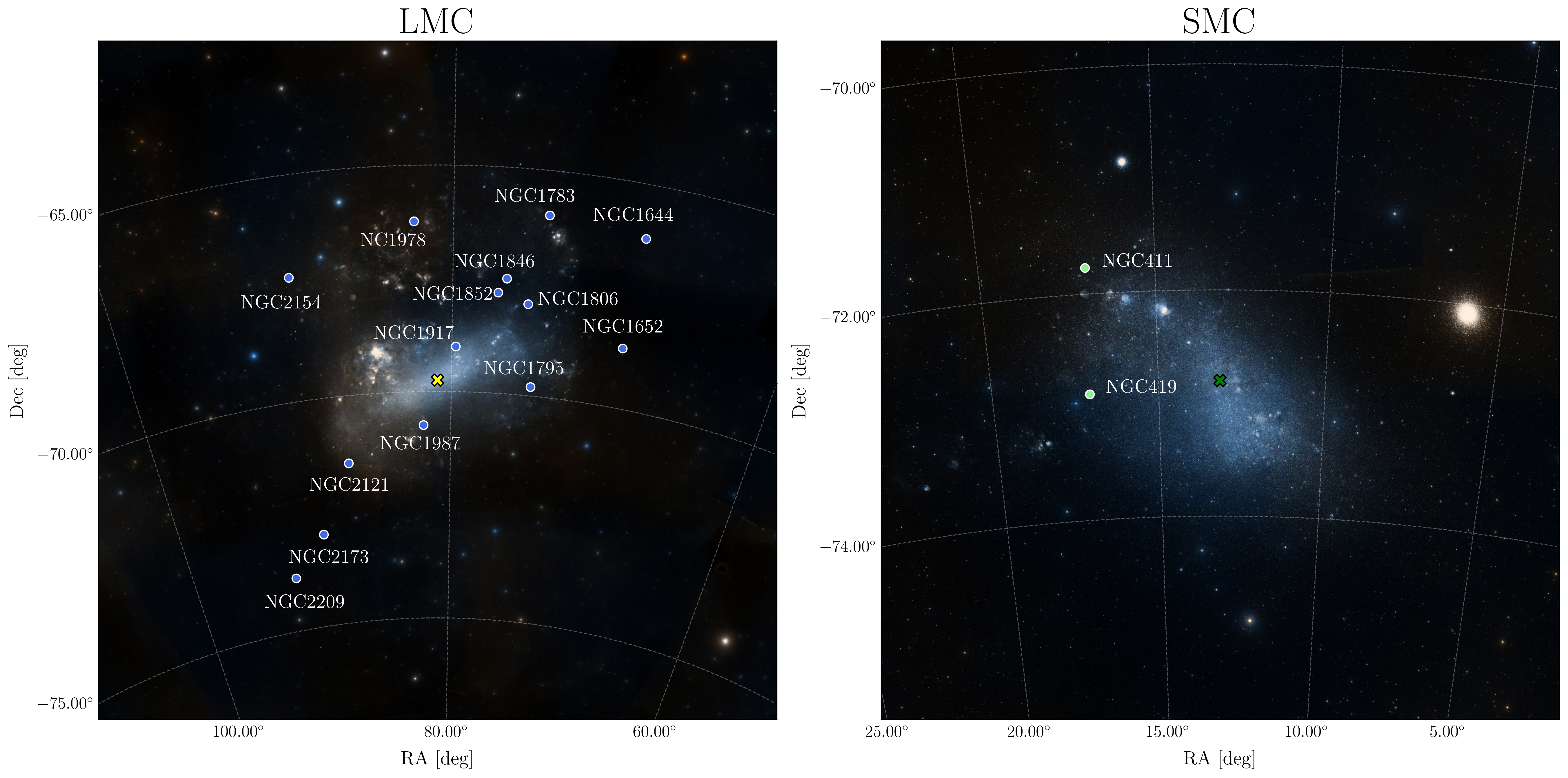}
    \caption{Image of the LMC and SMC obtained from the Digitized Sky Survey 2. Blue and green dots show the position of all the targets around the LMC and SMC, respectively. The yellow and green crosses mark the
adopted center of the LMC and SMC \citep{makarov2023, piatti2021}. North is up, and east is to the left.}
    \label{projection}
\end{figure*}

In this study, we analyzed 16 MC clusters spanning ages from ~1.3 to 2.9 Gyr and masses from $10^4$ to $10^5$ $M_{\odot}$.
Figure \ref{projection} presents the projected locations of all targets, comprising 14 clusters in the LMC and 2 in the SMC.

To investigate binary systems and low-mass stellar populations in our target clusters, we used the F435W, F555W, and F814W catalogs from \cite{milone2023}.
These measurements were obtained with the Wide Field Channel of the Ultraviolet and Visible Camera \citep[UVIS/WFC3;][]{mackenty2010} and with the Wide Field Channel of the Advanced Camera for Surveys \citep[ACS/WFC;][]{ford2003} onboard HST. The observations used for the catalogs are listed in Tab. B.1 of \cite{milone2023}.
As explained in Sect. \ref{sec_bin}, we restricted our analysis to these three filters because the CMD built from them maximizes the separation between single and binary stars, making them crucial for our objectives.

Stellar photometry and astrometry from \cite{milone2023} were performed with Jay Anderson's KS2 software package \citep{anderson2008}, which analyzes all exposures simultaneously using three complementary methods tailored to different brightness regimes \citep[e.g.\,][]{sabbi2016a, bellini2017a, milone2023}. \textit{Method I} detects sources through a 5$\times$5 pixel peak search, then fits a PSF \citep{anderson2000} to measure flux and position in each exposure, with the sky estimated in a 4--8 pixel annulus. \textit{Methods II and III} target stars too faint for reliable PSF fitting and instead use aperture photometry after neighbor subtraction: Method II applies weighted photometry in a 5$\times$5 pixel box, while Method III uses a 0.75-pixel circular aperture with a 2--4 pixel annulus, optimized for extremely crowded fields. To ensure high-precision measurements, we retained only isolated stars well described by the PSF model \citep[Sect.\,2.4]{milone2023}. Photometry was calibrated to the Vega system using encircled-energy corrections and STScI zero points \citep{milone2023}\footnote{ACS: \url{https://www.stsci.edu/hst/instrumentation/acs/data-analysis/zeropoints}; UVIS/WFC3: \url{https://www.stsci.edu/hst/instrumentation/wfc3/data-analysis/photometric-calibration}.}, and coordinates were corrected for geometric distortion following \citet{anderson2022a}\footnote{\url{https://www.stsci.edu/stsci-research/research-directory/jay-anderson}}.

For each cluster, we performed artificial star (AS) tests to evaluate the photometric uncertainties and to construct the simulated CMDs. Using the approach described by \cite{anderson2008}, we generated a catalog of 100,000 ASs, configured to replicate both the radial distributions and the luminosity functions of the observed stars. 
The instrumental magnitudes of the ASs range from about $-$13.6 mag to $-$9.0 mag in the ACS/WFC F814W filter. For the other filters, we assign magnitudes following the fiducial MS line, derived by linearly interpolating the median colors and magnitudes of the selected samples of stars over 0.5 mag wide F814W bins.
To measure the magnitudes and positions of the ASs, we used the KS2 software, adopting the same procedure as \cite{milone2023} used for real stars. Our analysis was limited to relatively isolated ASs that were well fitted by the point-spread function and that satisfied the same selection criteria applied to the real stars. These ASs are then used to estimate the completeness of each target.

The completeness depends on both the degree of crowding and the stellar luminosity. We followed the procedure described in \cite{milone2009} where ASs are employed to investigate how incompleteness and photometric uncertainties change as a function of crowding and brightness. For each cluster, we divided the field into concentric annuli centered on the cluster core, and within each annulus, we analyzed the AS outcomes in eight magnitude intervals, spanning -13.6 to -9 mag. The mean completeness in each bin was computed as the ratio between the number of recovered artificial stars and the number originally injected. This two-dimensional grid then enabled us to infer the completeness for any given star at any location within the cluster. 
To assess the robustness of our completeness estimates, we compared results obtained from different AS samples. The completeness fractions differ by less than 1\% between runs with 150,000 and 50,000 stars, supporting our adopted number of AS.
Finally, we interpolated the grid points and derived the completeness value associated with each star.

\section{Density profile}

Reliable determinations of the structural parameters of star clusters are necessary to describe their binary populations. For instance, the binary fraction within the core radius is crucial for comparing clusters that have different properties. 
Considering the distance of those clusters, the HST field of view also includes field stars projected along the same line of sight as each cluster, allowing us to study the density profile of each cluster.

\begin{figure}
    \centering
    \includegraphics[width=1.\linewidth]{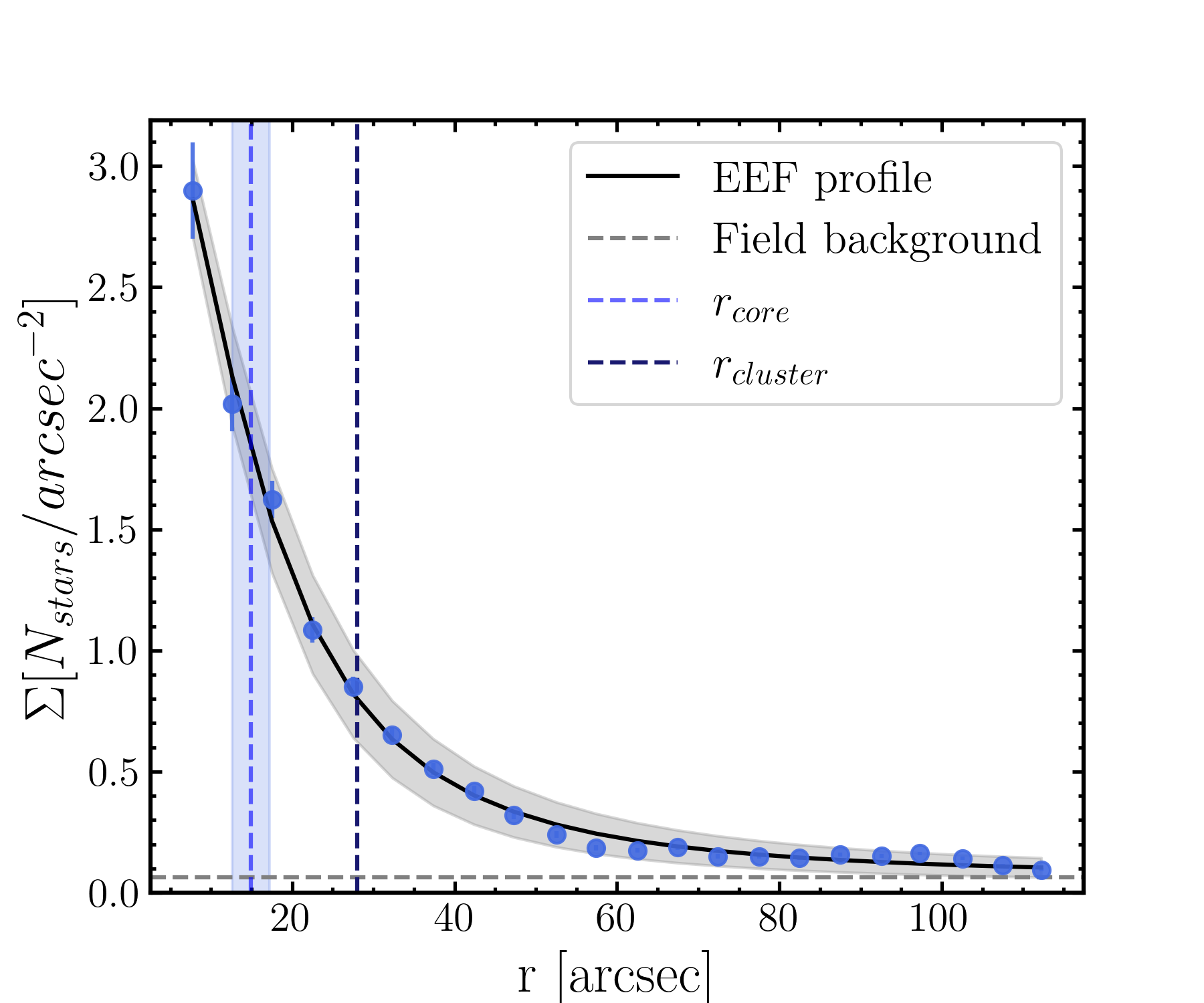}
    \caption{Density profile of NGC\,1806. The blue points correspond to the observed density profile of the cluster NGC\,1806, the contribution from field contamination is denoted by the grey dashed horizontal line, the black line is the best-fit EEF profile, and the vertical blue lines correspond to the core and the cluster radii, respectively. The shaded areas correspond to the 1-$\sigma$ dispersion.}
    \label{profile}
\end{figure}

First, based on the positions of all stars within the field of view, we identify the region that minimizes contamination from both background and foreground stars, hereafter referred to as the cluster field.
Following the procedure described by \citet[][see their Sect. 3.1]{mohandasan2024}, we divided the field of view into overlapping annular bins of constant width, defined as the difference between the external and internal radius. The width has been chosen to maximize spatial information while ensuring good statistics. We calculated the stellar density in each annulus ($\Sigma$), as the number of stars per square arcsec.
We restricted our analysis to stars within a magnitude range for which the completeness in the central region exceeds 0.5, thereby excluding stars significantly affected by incompleteness. This represents a refinement of the procedure adopted by \citet{mohandasan2024}, from which some differences arise. For NGC\,1806, which we use as a test case, this restriction corresponds to $m_{F814}<22.5$.
We plotted in Fig. \ref{profile} $\Sigma$ as a function of the average radius of each annulus (r) and performed a least-square fit of the observed density profile with the function discussed in \citet[][EEF profile]{elson1987}:

\begin{equation}
    \mu(r) = \mu_0 \left( 1 + \frac{r^2}{a^2} \right)^{-\gamma/2}+bg
    \label{eq:mu_r}
\end{equation}
where $\mu_0$ represents the central density in stars $arcsec^{-2}$, $ a$ is the scale radius, and $\gamma$ is the index of the power law at large radial distances from the cluster center, and $bg$ is a constant for the background counts. The cluster core radius, $r_{core}$, is linked to a, $\gamma$ by the relation:

\begin{equation}
    r_{core} = a (2^{2/\gamma} - 1)^{1/2}
    \label{eqrc}
\end{equation}
The error bars in each bin, which account for the Poisson noise of the observed star counts and the uncertainty in the completeness correction itself, are estimated as $\sigma_{N_i}= \sqrt{\frac{N_i}{C^2_i} +\frac{N^2_i}{ C^3_i}\frac{ 1-C_i}{N_{AS,i}} }$ 
where $C_i$ is the average completeness correction for the stars in the bin and $N_{AS,i}$ the number of injected ASs.
In summary, we calculated the number density of stars that are brighter than the magnitude at which the completeness drops by 0.5 in the innermost annulus, and evaluated this quantity at various radial distances from the cluster center. 
The profile with the EEF fit is shown in Fig. \ref{profile}.
Additionally, from the density profile we derived the $r_{50}$, which encloses half of the stars.
We defined by visually inspection $r_{cluster}$ as the radius within which the star counts are predominantly due to cluster members, and we used the most distant portion of the HST field of view, where the counts are dominated by field stars, to define $r_{field}$, i.e., the field region. 
To ensure that our choice does not influence the binary fraction or the MF slope, we repeated the analysis using different radial cuts within a range around the adopted radius. We confirmed that changing the separation radius does not substantially alter either the inferred binary fraction or the MF slope. 
The radii adopted for the cluster and field regions, within which we selected cluster members and field stars, respectively, are reported in Table\,\ref{tab2} for each cluster.

\begin{table*} 
    \centering
    \caption{Parameters of target clusters. }
    \begin{tabular}{cccccccccl}
        \hline
        Name    & $age\,\, $  & $E(B-V)$  & $(m-M)_0$& [M/H] &$\mu_0$& $r_{core}$  & $r_{50}$  & $r_{field}$  &$r_{cluster}$  
\\
         & [Gyr] & $[mag]$ & $[mag]$ & [dex] & $[arcsec^{-2}]$& [arcsec] & [arcsec] & [arcsec]
         &[arcsec] 
\\ \hline
        NGC\,411  & 1.65 & 18.97 & 0.06 & -0.7 &  2.64$\pm$0.09  & 11.62$\pm$1.40& 15.50$\pm$0.08& 93.6   &24.0   
\\ 
        NGC\,419  & 1.80 & 18.85 & 0.07 & -0.7 &  17.55$\pm$0.19 & 10.80$\pm$0.42& 22.80$\pm$0.04& 48.0  &28.0   
\\
        NGC\,1644 & 1.95 & 18.45 & 0.05 & -0.6 &  2.10$\pm$0.07   & 8.50$\pm$0.92& 11.60$\pm$0.02& 113.0 &28.0   
\\ 
        NGC\,1652 & 2.25 & 18.45 & 0.08 & -0.6 &  1.14$\pm$0.06  & 11.13$\pm$2.50& 11.70$\pm$0.10& 106.3  &20.0   
\\ 
        NGC\,1783 & 1.65 & 18.52 & 0.07 & -0.4 &  4.70$\pm$0.14   & 19.50$\pm$2.79& 27.50$\pm$0.22& 145.7  &32.0   
\\ 
        NGC\,1795 & 1.70 & 18.45 &  0.09 & -0.4 &  1.00$\pm$0.05   & 14.50$\pm$4.99& 25.40$\pm$0.02& 148.1  &20.0   
\\ 
        NGC\,1806 & 1.60 & 18.52 & 0.04 & -0.4 &  3.54$\pm$0.14  & 14.90$\pm$2.31& 21.10$\pm$0.14& 91.4 &28.0   
\\ 
        NGC\,1846 & 1.60 & 18.54 & 0.05 & -0.4 &  1.88$\pm$0.06  & 24.00$\pm$5.27& 27.80$\pm$0.17& 88.7 &28.0   
\\ 
        NGC\,1852 & 1.75 & 18.52 & 0.07 & -0.4 &  1.89$\pm$0.08  & 16.30$\pm$3.98& 25.50$\pm$0.07& 78.5  &28.0   
\\ 
        NGC\,1917 & 1.70 & 18.35 & 0.05 & -0.3 &  1.51$\pm$0.09  & 17.70$\pm$7.72& 27.00$\pm$0.06& 108.5  &20.0   
\\ 
        NGC\,1978 & 2.50 & 18.53 & 0.07 & -0.5 &  12.03$\pm$0.29 & 12.44$\pm$1.13 & 24.11$\pm$0.13 & 84.6  &24.0   
\\
        NGC\,1987 & 1.35 & 18.35 & 0.07 & -0.6 &  1.92$\pm$0.15  & 12.10$\pm$3.12 & 20.30$\pm$0.18& 123.7  &28.0   
\\
        NGC\,2121 & 2.90 & 18.42 & 0.14 & -0.5 &  3.57$\pm$0.27  & 18.45$\pm$4.83 & 28.02$\pm$0.36 & 88.0  &32.0   
\\
        NGC\,2154 & 2.00 & 18.43 & 0.04 & -0.5 &  2.32$\pm$0.07  & 14.74$\pm$2.22 & 20.63$\pm$0.06 & 104.7  &28.0   
\\ 
        NGC\,2173 & 2.05 & 18.37 & 0.06 & -0.4 &  1.90$\pm$0.04   & 14.02$\pm$1.41 & 16.50$\pm$0.06  & 98.9  &28.0   
\\
        NGC\,2209 & 1.45 & 18.34 & 0.15 & -0.4 &  0.76$\pm$0.08  & 17.87$\pm$6.80 & 58.88$\pm$0.07 & 120.5  &28.0   \\ \hline
    \end{tabular}
 \tablefoot{Ages, E(B-V), (m-M), and [M/H] are from isochrone fitting. $\mu_0$, $r_{core}$, and $r_{50}$ were evaluated from the profiles; $r_{cluster}$ was estimated by eye from the positions of cluster stars.}
    \label{tab1}
\end{table*}

\vspace{-0.5cm}

\section{Binary fraction through the Binary Map }
\label{sec_bin}
To investigate the population of binaries along the main sequences (MSs) of young clusters of MC, we used the \textit{binary map}, a new diagnostic tool introduced by \cite{muratore2026}, designed to distinguish between binaries with different mass ratios.

\begin{figure}
    \centering
    \includegraphics[width=1.\linewidth]{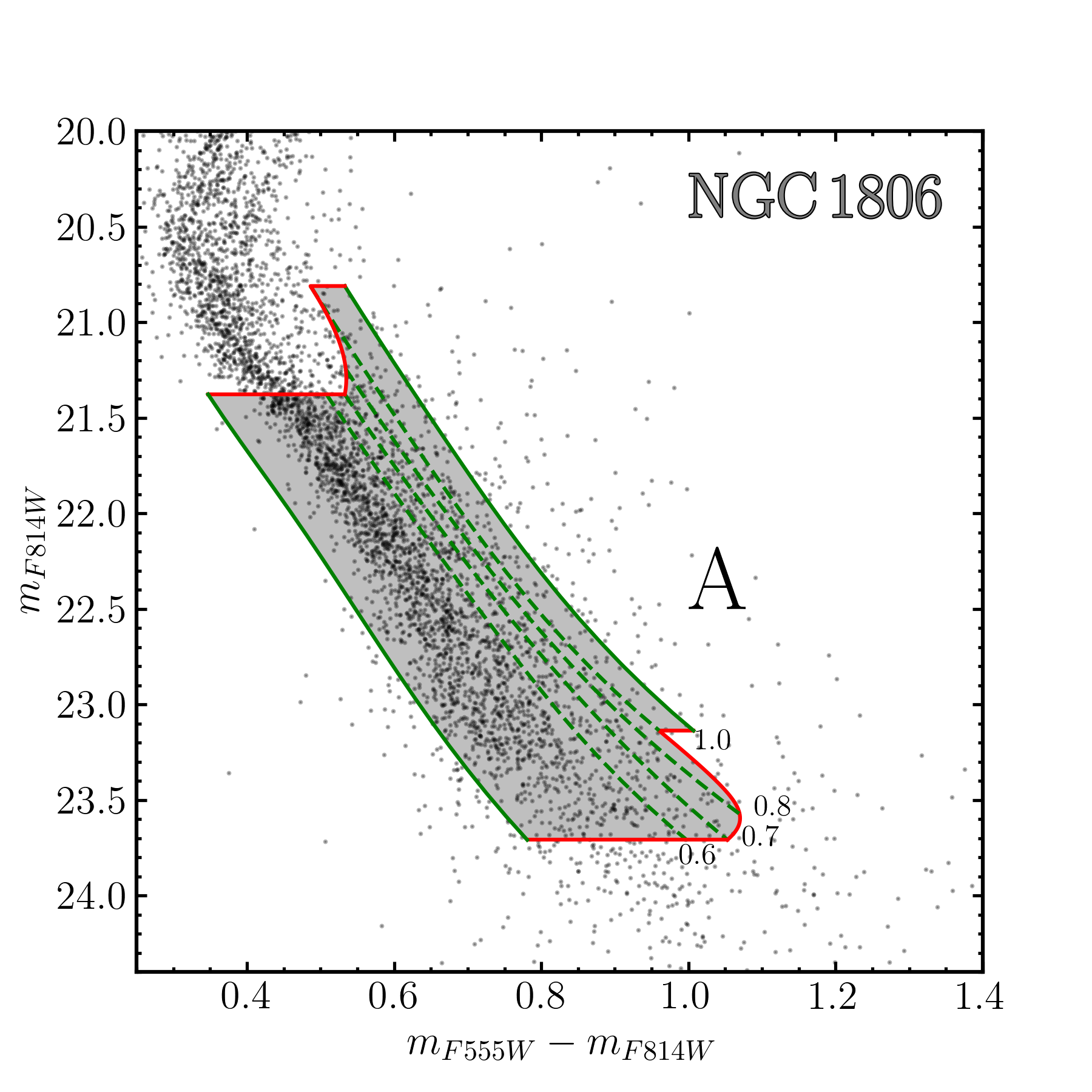}
\caption{
$m_{\rm F555W}$ vs.\,$m_{\rm F814W}-m_{\rm F814W}$ CMD of NGC\,1806, where the grey shaded area indicates the region used to study binaries and derive the \textit{Binary Map}. The six green lines delimit the five subregions (R1--R5) adopted for the binary analysis. The bluest line represents the fiducial sequence of single stars, shifted to the blue to encompass the bulk of MS stars, while the reddest line corresponds to the fiducial of equal-mass binaries, shifted to the red to include most binaries. The remaining lines represent fiducial sequences of binaries with mass ratios ranging from 0.6 to 1.0, as indicated.
}

    \label{cmd_strips}
\end{figure}

This diagram is based on the well-known photometric properties of unresolved binary systems composed of two MS stars, which appear as single point-like sources with a total magnitude

\begin{equation}
    m_{\mathrm{bin}} = m_{1} - 2.5 \log \left( 1 + \frac{F_{2}}{F_{1}} \right),
\end{equation}

where $m_{1}$ is the magnitude of the primary (brightest) star and $F_{1}$ and $F_{2}$ are the fluxes of the primary and secondary components, respectively. 
In a simple stellar population, the flux of MS stars depends on their mass following a specific mass-luminosity relation. As a consequence, the total magnitude of a binary system composed of two MS stars can be uniquely determined by the mass of the primary component and the mass ratio, $q = M_{2} / M_{1}$, where $M_{1}$ and $M_{2}$ are the masses of the primary and secondary stars. 
To construct the \textit{Binary Map}, we used the setup illustrated in Fig.\,\ref{cmd_strips}, where the shaded area superimposed on the CMD of NGC\,1806 highlights the region, hereafter referred to as region~A, used to study binary systems. This region comprises the MS segment where binaries with a mass ratio larger than 0.7 are well separated from single MS stars. 
In the test case NGC\,1806, this region includes single stars with $20.8 < m_{\rm F814W} < 23.8$ and binary systems whose primary component falls within the same magnitude interval.

As shown in Fig.\,\ref{cmd_strips}, we defined six green reference lines, hereafter referred to as lines~1--6 (from the bluest to the reddest in color). 
Line~1, which represents the left boundary of region~A, corresponds to the fiducial line of MS shifted toward the blue by two times the average color error ($\sim$0.05\,mag) in order to enclose the bulk of MS stars. 
Lines~2 to~5, the dashed ones, correspond to the fiducial sequences of binary systems with mass ratios $q = 0.6$, 0.7, 0.8, and 1.0, respectively. These fiducial lines were derived following the same procedure adopted in our previous works \citep[e.g.][]{milone2012,muratore2024}, using the MS fiducial line and the mass--luminosity relation provided by the best-fit isochrone to compute the colors and magnitudes corresponding to binaries with different $q$ values. 
Finally, line~6 represents the fiducial sequence of equal-mass binaries shifted toward the red by two times the average color error ($\sim$0.05\,mag), to include binaries with large mass ratios that appear redder due to observational uncertainties.
These six lines define five regions, namely $R_1$--$R_5$. Region $R_1$ includes the bulk of single stars and binaries with a mass ratio smaller than 0.5. 
Regions~$R_2$--$R_4$ contain binaries in successive $0.1$-wide $q$ bins from $0.6$ to $0.8$, while region~$R_5$ covers $0.8\leq q\leq1.0$ (width $0.2$).
Region 6 primarily hosts binaries shifted to the red by photometric uncertainties. 

We applied a two-step verticalization. 
In the first step, we normalized the colors such that lines~1--6 are transformed into vertical lines. 
To achieve this, we adapted Eq.~1 from \citet{milone2025a} to each region ($R_{i}$, $i=1$--5) and derived the quantities

\begin{gather}
\delta_{F555W,F814W}^{i} =
\begin{cases}
     W_{i} \, \frac{X - X_{\mathrm{line}\,i}}{X_{\mathrm{line}\,i} - X_{\mathrm{line}\,i+1}} + \sum_{j=1}^{i-1} W_{j} \quad \text{if } X \in R_i \\
   \hspace{2cm}  0  \hspace{1.5cm} \quad \text{else }
\end{cases}
\end{gather}

where $X = m_{\mathrm{F555W}} - m_{\mathrm{F814W}}$, and $W_{i}$ is the color separation between lines~$i+1$ and~$i$, measured at $m_{F814W}=$22.0\,mag. 
We then combined these quantities to define the pseudo-color:

\begin{equation}
    \Delta_{F555W,F814W}^{Bin} = \sum_{i=1}^{5} \delta_{F555W,F814W}^{i}.
\end{equation}

In the second step, the verticalization is performed in magnitude. Specifically, for each region, we define the quantity
\begin{equation}
\label{eq3}
    \Delta_{F814W} = W_{y} \, \frac{Y - Y_{\mathrm{top\,line}, i}}{Y_{\mathrm{top\,line}, i} - Y_{\mathrm{bottom\,line}, i}},
\end{equation}
where $Y = m_{\mathrm{F814W}}$, and $W_{y}$ denotes the magnitude interval of region~$R_1$. The top line corresponds to the red top line from $m_{\mathrm{F814W}} = 21.5$ in region~$R_1$, to the segment with $m_{\mathrm{F814W}} = 21.5 - 0.75$ in region~$R_5$, tracing binaries whose primary star has $m_{\mathrm{F814W}} = 21.5$ and mass ratios between 0.5 and~1.0 in the remaining regions. 
The red bottom line is defined analogously, but for $m_{\mathrm{F814W}} = 23.8$.

The resulting $\Delta_{F814W}$ versus $\Delta_{F555W,F814W}^{Bin}$ \textit{binary map} for NGC\,1806 is shown in the left panel of Fig.\,\ref{binary_map}, together with the $\Delta_{F555W,F814W}$ histogram. The majority of stars lie within the region~1--2, but a well-populated tail extends into the regions~3--5, confirming that NGC\,1806 hosts a significant fraction of binaries with $q > 0.6$.

\subsection{The fraction of binaries in MC young clusters}

To reproduce the observed stellar distribution across the \textit{binary map}, we adopted a procedure based on ASs described in \cite{muratore2026}. It works iteratively: at each step, ASs are injected into each region of the map. Since observational uncertainties can scatter stars from one region to another, the number of recovered ASs in a given region generally does not match the number injected in that region. The residual difference between the observed and recovered counts is then used to adjust the number of ASs to inject in the next iteration. The process is repeated until the number of recovered stars in every bin converges to a certain criterion, naturally accounting for both photometric scatter and completeness effects.
We set the convergence criterion 
 \[
 \big|\, n_{\mathrm{OBS},i} - N_i \,\big| < N_{FIELD,i}
 \]
 Where \(N_i\) represents the number of stars observed in region \(R_{i}\), \(n_{\mathrm{OBS},i}\) refers to the number of recovered ASs in each bin, while \(N_{FIELD,i}\) denotes the observed number of field stars. 
The value of \(N_{FIELD,i}\) is determined by using a sample of stars located in a region away (see $r_{field}$ in Tab. \ref{tab1}) from the cluster center, with an area that is the same as the area used for cluster membership analysis.

\begin{figure}
    \centering
    \includegraphics[width=.6\textwidth]{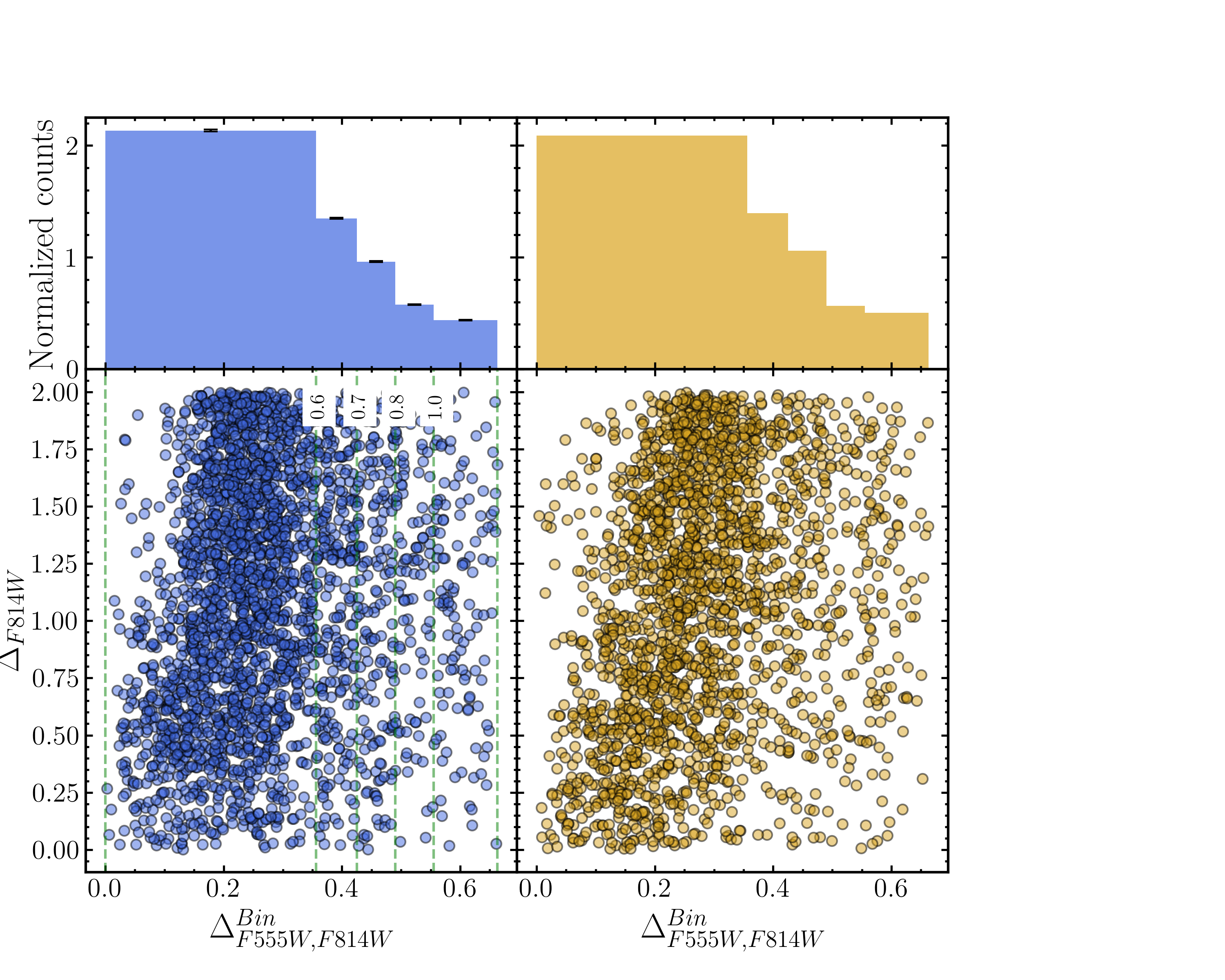}
    \caption{\textit{Binary map} for NGC\,1806. Left: the observed \textit{Binary map} (in blue), with the corresponding $\Delta_{F555W,F814W}$ distribution shown in the top panel. The dashed vertical lines indicate the six reference loci used to construct the map, including the fiducial sequences of binary systems with different mass ratios, $q$, whose values are labeled in the figure. Right: the best-fit \textit{Binary map} (in gold). 
    }
    \label{binary_map}
\end{figure}

Figure~\ref{binary_map} compares the observed \textit{binary map} of NGC\,1806 (left) with the simulated best-fit (right). The corresponding $\Delta_{\mathrm{F555W,F814W}}^{\mathrm{Bin}}$ histogram distributions are shown in the top panels.
This procedure provides the numbers of single stars and binaries with their respective mass ratios.

Thanks to this procedure, we determined the binary fractions for stars throughout the full $r_{cluster}$, within the $r_{core}$, and, where possible, in the region between the core and the R$_{50}$.
We consider only binaries with mass ratios greater than 0.7 because the small color separation makes it challenging to robustly distinguish low-mass-ratio binaries from single stars. 

By combining the results of the two bins of NGC\,1806, we derived a binary fraction with a mass ratio greater than 0.7 of $f_{\mathrm{bin}}^{q>0.7} = 0.07 \pm 0.01$. 
The uncertainty on the binary fraction is estimated using 1000 Monte Carlo simulations. In each simulation, we generated CMDs with a large number of stars, assuming a flat binary fraction that matches the observed one. Then, we randomly draw stars in quantities similar to those in the observed CMD and evaluate the binary fraction. The error is 1$\sigma$ of the fraction distribution arising from simulations.

Assuming a flat mass-ratio distribution, we infer an overall binary fraction of $f_{\mathrm{bin}}^{\mathrm{TOT}} = 0.24 \pm 0.02$, taking into account a minimum mass of 0.075 $M_{\odot}$ for the secondary star. 
Given the distance of the MC, the photometry does not allow us to determine the fraction in more than two bins in q when assessing the distribution. However, we were able to measure the ratio $R_q$ between the two inferred bins, providing an indication of the q distribution.  
For NGC\,1795, NGC\,1917, NGC\,1978, and NGC\,1987, contamination and photometric uncertainties are more significant than in the other clusters; therefore, we report results only for a single bin, spanning q=0.7--1.0.
The derived parameters are summarized in Table~\ref{tab2}.
The $R_q$ ratios for NGC\,411, NGC\,1806, NGC\,2121, NGC\,2154, NGC\,2173, and NGC\,2209 are close to $1$, consistent with a flat mass-ratio distribution. NGC\,419, NGC\,1644, NGC\,1652, and NGC\,1783 exhibit ratios greater than 1.35, suggesting a potentially negative slope in the mass-ratio distribution, whereas NGC\,1846 and NGC\,1852 have ratios below unity, implying a possibly positive slope. Nevertheless, the average of all ratios is 1.04 with a standard deviation of 0.31, suggesting a flat mass-ratio distribution.  

Additionally, our assumption of extrapolating the total binary fraction is supported by homogeneous studies of stellar populations in both Galactic and MC clusters \citep{milone2012, cordoni2023}. The impact of this assumption can be tested by adopting alternative mass-ratio distributions. Due to photometric errors, binaries with small mass ratios are indistinguishable from single MS stars in this dataset, so it is impossible to determine the total MS-MS binary fraction without assuming a specific mass ratio distribution. Following \cite{milone2012}, we compute the extrapolation factor for a flat (the one used in our extrapolation), a Fisher, and a random-pair distribution, obtaining 3.33, 2.32, and 7.14, respectively. As a consequence, when adopting a Fisher distribution that is picked higher mass ratio values, the extrapolated fraction decreases by a factor of 0.70, whereas, under the assumption of a random distribution picked toward lower mass ratios, we obtain an increased extrapolated fraction of 2.14.

\begin{table*}
    \centering
    \caption{Inferred parameter of target clusters. }
    \begin{tabular}{cccclcccc}
        \hline
        name & $f_{bin}^{q\ge0.7}$& $f_{bin,r_{core}}^{q\ge0.7}$ & $f_{bin,r_{50}}^{q\ge0.7}$  &$f_{bin}^{ext}$& $R_q$ & $f_{BS}$ & $f_{BS,r_{core}}$ & $\alpha$ \\
        \hline
        NGC\,411 & $0.06\pm0.01$  & $0.07\pm0.02$ & $0.10\pm0.02$ & $0.19\pm0.03$  & $0.90\pm0.35$ & $0.040\pm0.009$ & $0.064\pm0.019$ & $-1.52\pm0.72$\\
        NGC\,419 & $0.07\pm0.01$  & $0.14\pm0.02$ & $0.07\pm0.02$ & $0.23\pm0.02$  & $1.28\pm0.22$ & $0.006\pm0.002$ & $0.010\pm0.004$ & $-1.92\pm0.17$\\
        NGC\,1644 & $0.10\pm0.01$ &  $0.18\pm0.03$ & -  &  $0.34\pm0.04$  & $1.38\pm0.40$ & $0.017\pm0.006$ & $0.015\pm0.008$ & $-2.48\pm0.20$\\
        NGC\,1652 & $0.06\pm0.01$ & $0.05\pm0.02$ & $0.05\pm0.02$  & $0.19\pm0.03$  & $1.38\pm0.53$ &  $0.026\pm0.012$ & $0.056\pm0.020$ & $-1.63\pm0.32$\\
        NGC\,1783 & $0.08\pm0.01$ & $0.06\pm0.01$ & $0.05\pm0.01$ & $0.27\pm0.02$  & $1.56\pm0.25$ & $0.013\pm0.003$ & $0.014\pm0.004$ & $-0.97\pm0.13$\\
        NGC\,1795 & $0.11\pm0.01$ & $0.10\pm0.02$ & $0.08\pm0.02$ & $0.36\pm0.05$  & - & $0.013\pm0.006$ & $0.013\pm0.090$ & $-1.95\pm0.27$\\
        NGC\,1806 & $0.07\pm0.01$ & $0.07\pm0.01$ & $0.07\pm0.01$ & $0.24\pm0.02$  & $1.04\pm0.18$ & $0.012\pm0.003$ & $0.017\pm0.006$ & $-1.20\pm0.20$\\
        NGC\,1846 & $0.09\pm0.01$ & $0.07\pm0.02$ & $0.07\pm0.02$  & $0.29\pm0.02$  & $0.51\pm0.12$ & $0.032\pm0.005$ & $0.032\pm0.005$ & $-0.92\pm0.14$\\
        NGC\,1852 & $0.11\pm0.01$ & $0.16\pm0.04$ & $0.10\pm0.02$  & $0.36\pm0.05$  & $0.62\pm0.24$ & $0.0\pm0.0$ & $0.0\pm0.0$ & $-1.53\pm0.46$\\
        NGC\,1917 & $0.06\pm0.01$ & $0.06\pm0.01$ & -  & $0.21\pm0.04$  &  - & $0.007\pm0.004$ & $0.019\pm0.008$ & $-1.69\pm0.28$\\
        NGC\,1978 & $0.06\pm0.01$ & $0.10\pm0.02$ & $0.05\pm0.01$  & $0.19\pm0.02$  & - & $0.014\pm0.002$ & $0.012\pm0.004$ & $-1.96\pm0.13$\\
        NGC\,1987 & $0.10\pm0.01$ & $0.12\pm0.02$ & $0.10\pm0.02$  & $0.34\pm0.04$  & - & $0.019\pm0.008$ & $0.014\pm0.007$ & $-2.53\pm0.12$\\
        NGC\,2121 & $0.05\pm0.01$ & $0.05\pm0.01$ & $0.05\pm0.01$  & $0.18\pm0.03$  & $0.82\pm0.21$ & $0.0\pm0.0$ & $0.013\pm0.006$ & $-0.83\pm0.44$\\
        NGC\,2154 & $0.08\pm0.01$ & $0.18\pm0.02$ & $0.07\pm0.02$  & $0.27\pm0.03$  & $0.97\pm0.24$ & $0.026\pm0.007$ & $0.033\pm0.012$ & $-1.37\pm0.15$\\
        NGC\,2173 & $0.13\pm0.02$ & $0.15\pm0.03$ &  -  & $0.45\pm0.05$  & $1.35\pm0.35$ & $0.043\pm0.010$ & $0.061\pm0.016$ & $-1.46\pm0.37$\\
        NGC\,2209 & $0.11\pm0.01$ & $0.13\pm0.02$ & -  & $0.37\pm0.04$  & $1.05\pm0.31$ & $0.018\pm0.009$ & $0.030\pm0.019$ & $-1.25\pm0.18$\\
        \hline
    \end{tabular}
    \tablefoot{Fractions of binaries within the cluster radius, core radius, and between the core radius and the R$_{50}$, the extrapolated fraction of binaries within the cluster radius; the ratio $R_q$ between the two bins in the q distribution, fraction of BSs within the cluster and core radius.}
    \label{tab2}
\end{table*}

\begin{figure*}
    \centering
    \includegraphics[width=1.\linewidth]{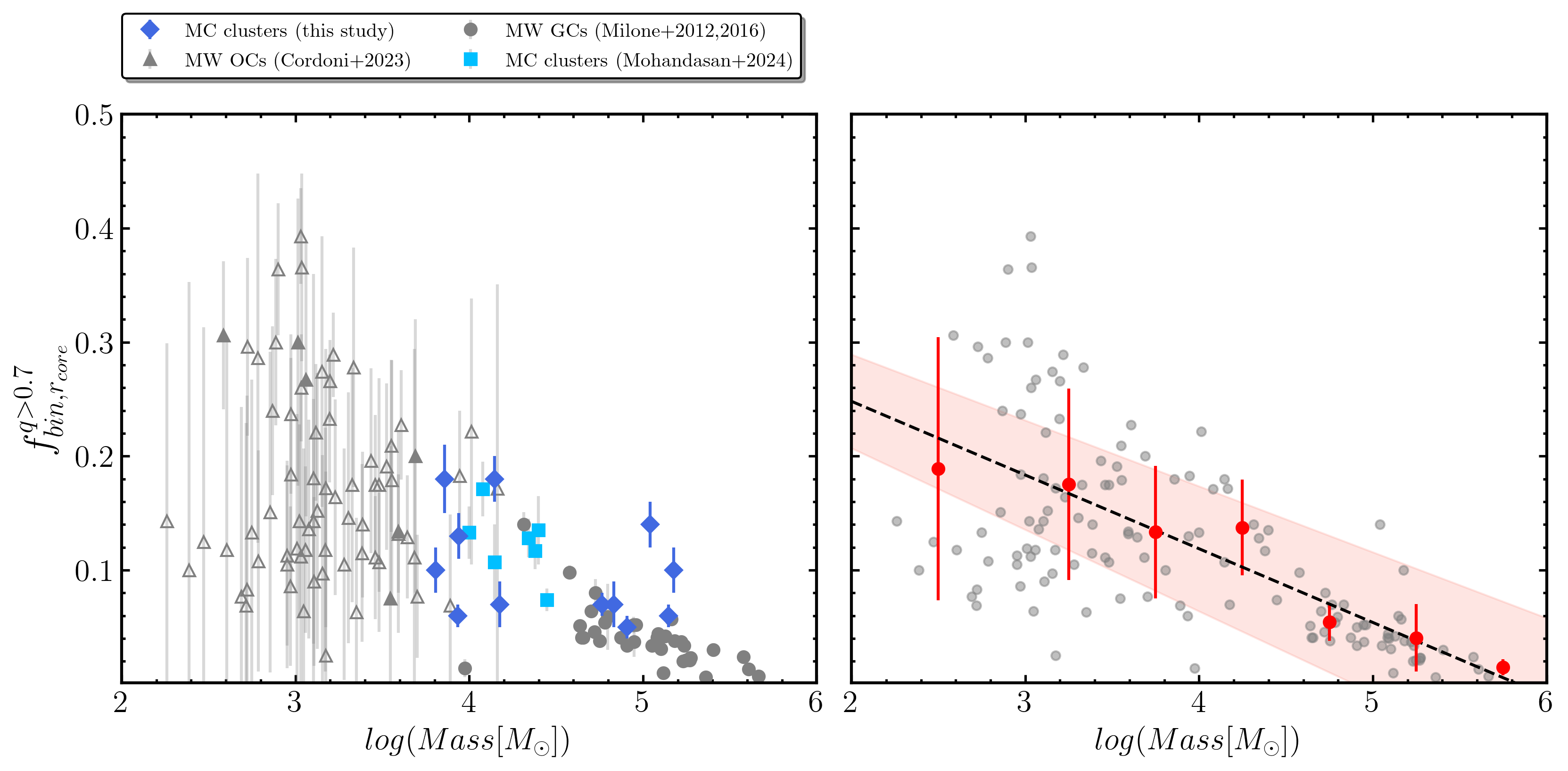}
    \caption{The fraction of binary systems within the core radius as a function of cluster mass over a broad range of values, encompassing both GCs \citep{milone2012, milone2016} and young stellar clusters in the Milky Way and MC \citep{cordoni2023, mohandasan2024}. The unfilled dots represent those clusters for which the fractions were extrapolated. In the right panel, the red dots indicate the average values of the binary fractions, while the red line and shaded region show the best-fit relation and its 1-sigma uncertainty range. The masses for GCs are taken from \cite{baumgardt2018}}
    \label{comparing_logmass}
\end{figure*}

\begin{figure*}
    \centering
    \includegraphics[width=1.\linewidth]{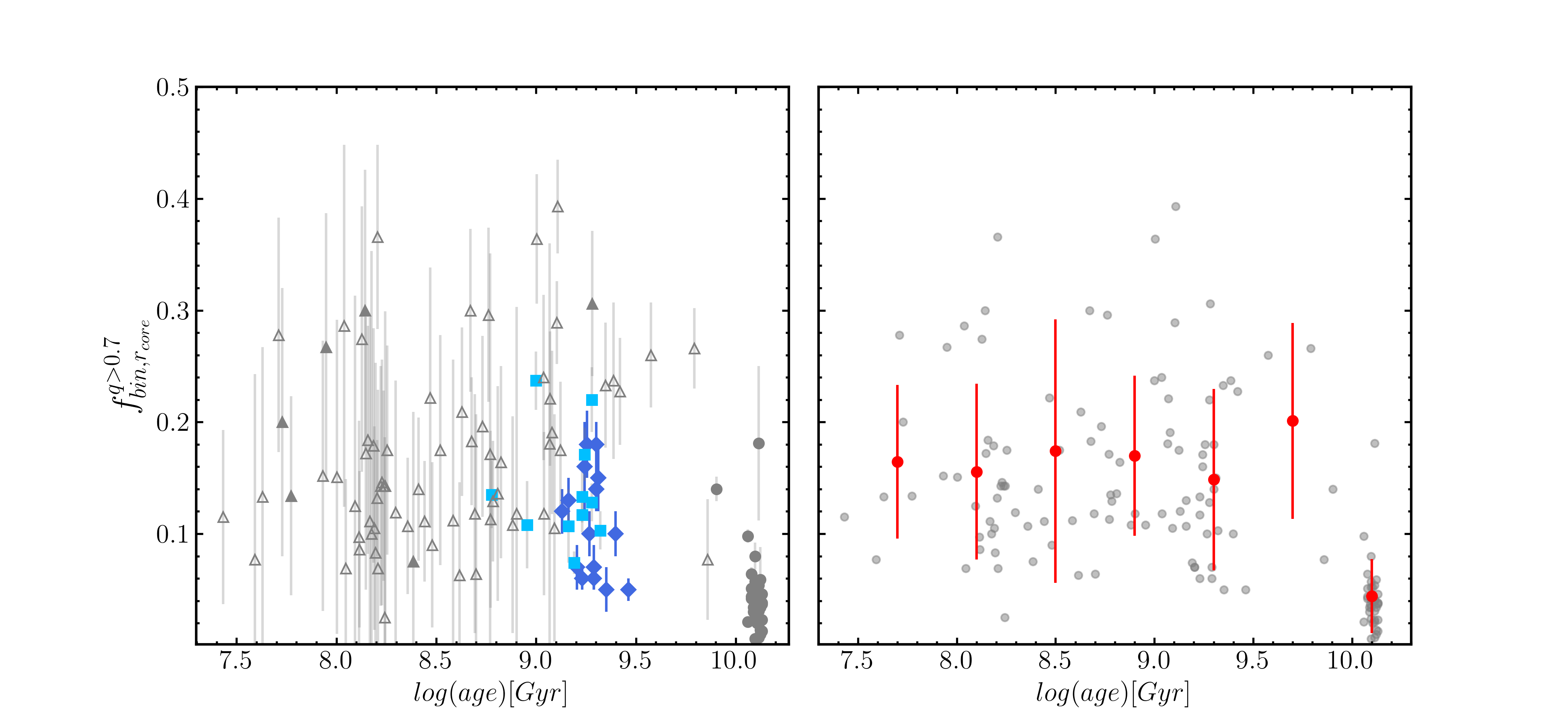}
    \caption{The fraction of binary systems within the core radius as a function of cluster age over a broad range of values, encompassing both GCs \citep{milone2012, milone2016} and young stellar clusters in the Milky Way and Magellanic Clouds \citep{cordoni2023, mohandasan2024}. The unfilled dots represent those clusters for which the fractions were extrapolated. In the right panel, the red dots indicate the average values of the binary fractions.}
    \label{comparing_logage}
\end{figure*}

\subsection{Comparison with other surveys}
To compare the binary fractions in MC star clusters with those in Galactic open and GCs, we merged our results with published studies. 
Taking into account all clusters together as a single sample, we confirm the anticorrelation between the binary fraction of the core and the mass of the cluster, as shown in Fig. \ref{comparing_logmass} and previously identified in \cite{milone2012,milone2016,cordoni2023, mohandasan2024}. 
To obtain the anticorrelation, we calculated the average binary fraction across all available data, within different mass bins. We then used these mean values for a linear fitting, resulting in a slope of -0.04 and an intercept of 0.28.
In particular, the MC clusters analyzed here display a binary fraction higher than most Galactic GCs of comparable mass.
All MC clusters fall within the range spanned by MW clusters, indicating there is no apparent dependence on environmental conditions.
On the other hand, as illustrated in the right panel of Fig. \ref{comparing_logage}, we find no indication of a correlation between the binary fraction and cluster age.

\section{Luminosity and Mass functions}
\begin{figure*}
    \centering
    \includegraphics[width=1.\linewidth]{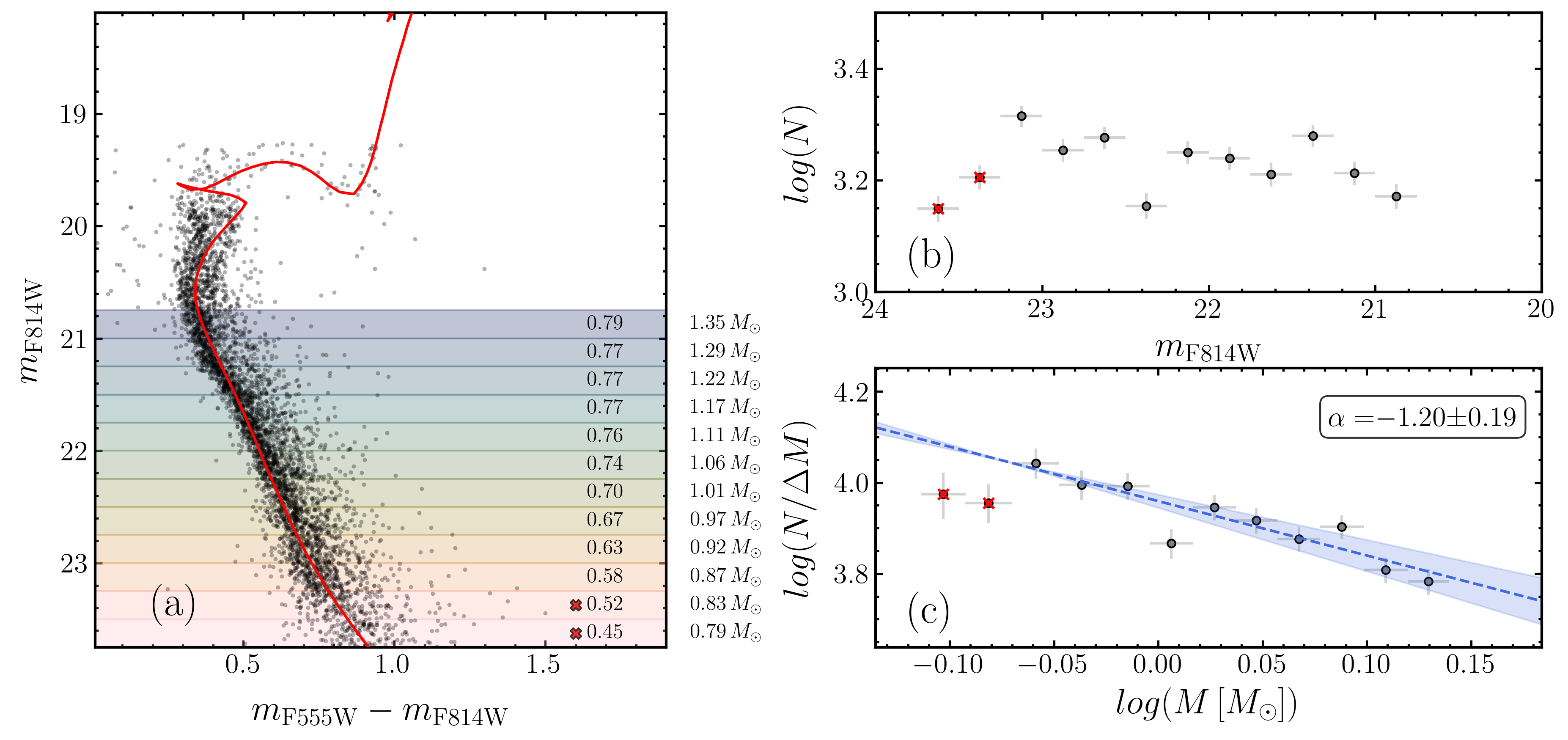}
    \caption{Determination of the MF of NGC\,1806. \textit{Panel a.} $m_{\rm F814W}$ vs.\,$m_{\rm F555W}-m_{\rm F814W}$ CMD of NGC\,1806 stars. The red solid line represents the Padova isochrone adopted to derive the mass–luminosity relation over the magnitude interval $20.8<m_{\rm F814W}<24$. The stellar masses associated with the magnitude bins used to derive the MF are on the right side of the panel. The shaded regions mark the bins adopted to derive the MF along the upper MS. The completeness relative to those bins is reported with the bin on the right, while the red crosses mark those bins with a low completeness. {\it Panels b and c.} $m_{\rm F814W}$ luminosity function (b) and corresponding MF (c). In panel c, the black solid line represents a linear fit to the observed MF, restricted to stars in the magnitude range with an accepted completeness.}
    \label{MF}
\end{figure*}

To derive the luminosity function of stars in the target clusters, we adapted the method introduced by \cite{milone2012b}. Panels a) of Fig.~\ref{MF} show the binning scheme used for the analysis; in particular, we excluded from our analysis turn-off stars which are in a fast evolutionary phase and are affected by differences in rotational rates \citep{milone2015a}.
We corrected the luminosity function for the contribution of unresolved binaries following the method introduced in \cite{legnardi2025}, adopting the total binary fractions inferred in the previous Sect.
The resulting luminosity function, corrected for completeness and binaries, is shown in panel b of Fig.~\ref{MF}. Red crosses indicate bins with low completeness, so we excluded that point in the next calculations, and error bars represent Poisson uncertainties. The red line overlaid on the CMD in panel a of Fig.~\ref{MF} corresponds to the best-fitting isochrone of the Padova isochrones \citep{marigo2017}.
This fit uses an age, metallicity, distance modulus, and reddening from Tab. \ref{tab1}. Using this isochrone, we converted the observed luminosities into stellar masses, as illustrated in the same panel. The distinct mass intervals used to construct the MF are shown in different colors, with the average mass of stars in each bin labeled accordingly. In addition, the values in the plot represent the mean completeness for each bin. The resulting MF is shown in panel c of Fig.~\ref{MF}, where we plot the logarithm of the number of stars per unit mass ($\log(dN/dM)$), normalized by the bin width $\Delta M$, as a function of the logarithm of stellar mass. The present-day MF of NGC\,1806 can be described by a power-law:

\begin{equation}
\frac{dN}{dM}=k \cdot M^{-\alpha}
\end{equation}

\noindent where $k$ is a normalization constant and $\alpha$ is the slope. Taking the logarithm of both sides gives the linear form:

\begin{equation}
\log\left(\frac{dN}{dM}\right) = \log(k) - \alpha \cdot \log(M)
\end{equation}
The error bars in each bin, which account for the Poisson noise of the observed star counts and the uncertainty in the completeness correction itself, are estimated in the same way described for the density profile in Sect. 3.
We perform a least-squares linear fit to the observed $\log(dN/dM)$ versus $\log(M)$ distribution, obtaining a best-fitting slope of $\alpha$ (solid black line). The uncertainties relative to the slopes were derived from the covariance matrix of the fitted coefficients obtained in the least-squares solution. All slopes are listed in Table \ref{tab2}.

\begin{figure}
    \centering
    \includegraphics[width=1.\linewidth]{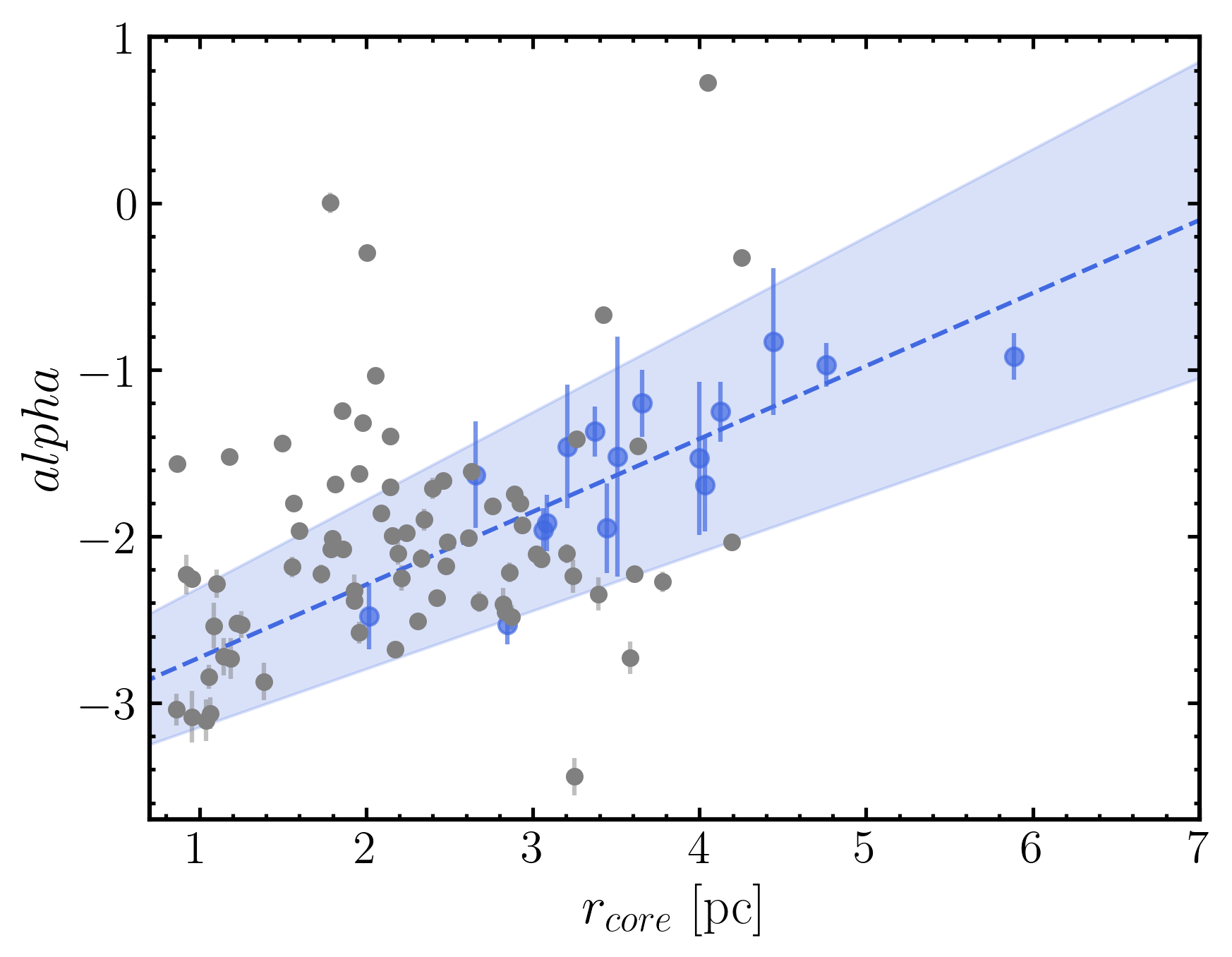}
    \caption{Core radius as a function of slopes of MFs. The clusters analyzed in this work are shown as blue dots, while the OCs from \cite{cordoni2023} are plotted in grey. The blue line and region show the best-fit relation and its 1-sigma uncertainty range for the target clusters. }
    \label{rcorealpha}
\end{figure}

Figure \ref{rcorealpha} illustrates how $r_{core}$ is related to the slope of the MF, in particular clusters with steeper slopes tend to have smaller core radii $r_{core}$. We fit with a linear function only the target clusters analyzed in this study, obtaining a slope of $0.44 \pm 0.09$. 
We computed the Pearson and Spearman coefficients, obtaining $r = 0.797$ and $\rho = 0.794$, both of which indicate a strong positive relationship between $R_{core}$ and $\alpha$. The very small p-values ($p \approx 2.2\times10^{-4}$ for Pearson and $p \approx 2.4\times10^{-4}$ for Spearman) show that this association is statistically significant in linear and monotonic terms.

\begin{figure*}
    \centering
    \includegraphics[width=1.\linewidth]{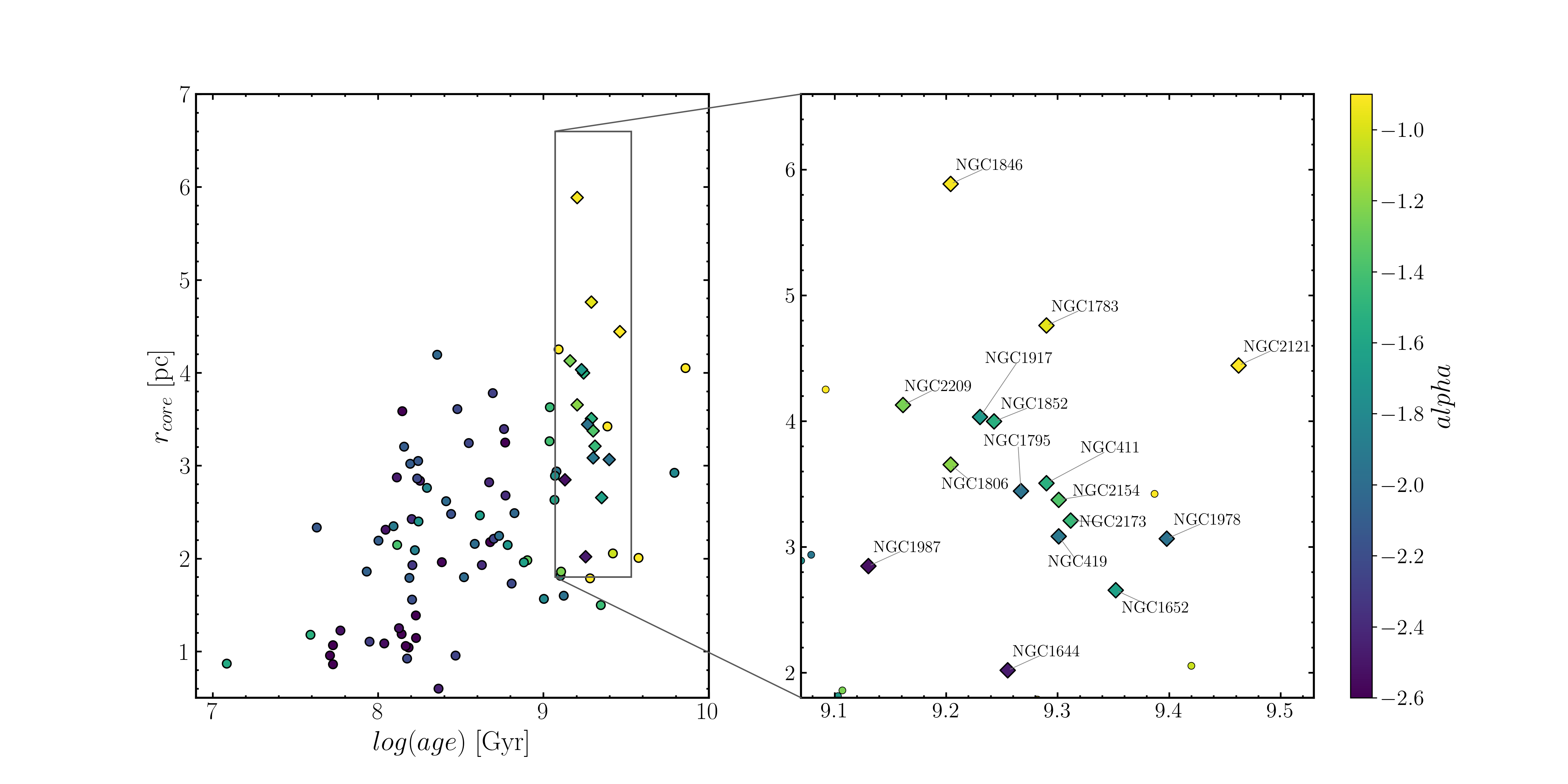}
    \caption{Core radius as a function of cluster age. In the left panel, the clusters analyzed in this work are shown as diamonds, while the OCs from \cite{cordoni2023} are plotted as circles, with colors indicating the slopes of their mass functions. The right panel provides a zoomed-in view of the region occupied by the clusters studied here.}
    \label{comparing1}
\end{figure*}

To have a broader perspective, Fig. \ref{comparing1} shows the log(age)–$r_{core}$ diagram for all target clusters and OCs of \cite{cordoni2023}, where the colors indicate the slopes of their MFs. 
The relation suggests that as clusters evolve dynamically and undergo expansion, their core radii are substantially altered, with this change related to the present-day MF. According to \cite{elson1991}, clusters characterized by a flatter IMF (i.e., containing a comparatively larger fraction of massive stars) experience more rapid mass loss due to stellar evolution, which drives faster and more significant core expansion, whereas clusters with a steeper IMF preserve dense and compact cores over much longer periods. Consequently, the structural evolution of a cluster is dependent on its IMF, and the present-day MF (the one evaluated in this study) encodes the cumulative effects of both the initial IMF and the subsequent dynamical evolution, including expansion. \cite{wilkinson2003} revisited this concept, testing two mechanisms for core expansion: the evolving tidal field of the LMC and a substantial primordial population of hard binaries. N-body simulations of clusters on circular and eccentric orbits around a point-mass LMC showed nearly identical core-radius evolution, implying the LMC tidal field has not yet significantly affected the structure of intermediate-age clusters. They further showed that although many primordial hard binaries can expand the core radius, the effect is too small to explain the full observed spread.

\section{Blue straggler stars }

\begin{figure}
    \centering
    \includegraphics[width=1.\linewidth]{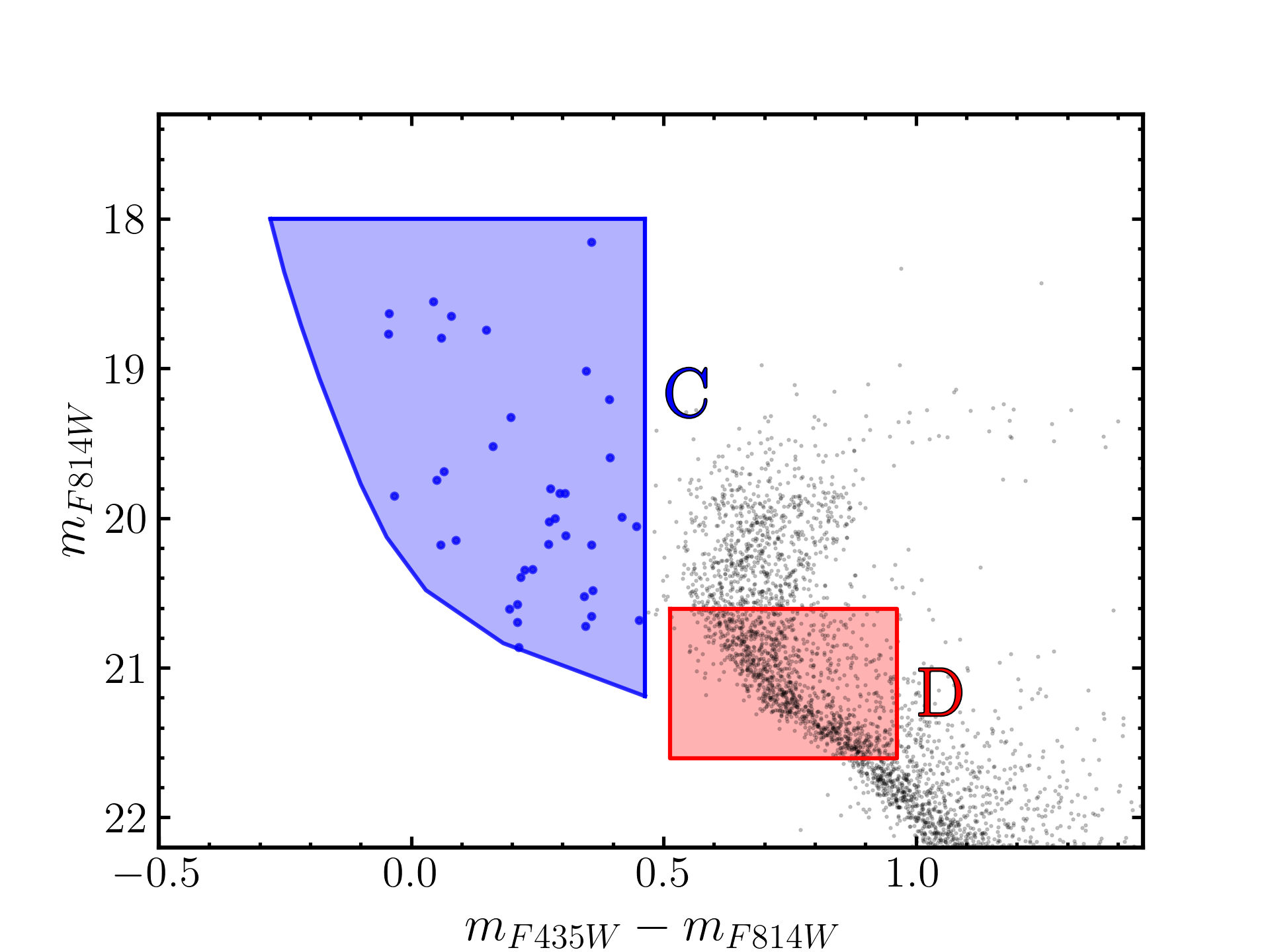}
    \caption{Region adopted to investigate the inference of BSs $m_{F814W}$ vs. $m_{F435W}-m_{F814W}$ CMD of NGC\,1806. The blue shaded area contains most of the BSs in the cluster, while the red shaded area is predominantly occupied by MS stars. }
    \label{BSs}
\end{figure}

\begin{figure}
    \centering
    \includegraphics[width=1.\linewidth]{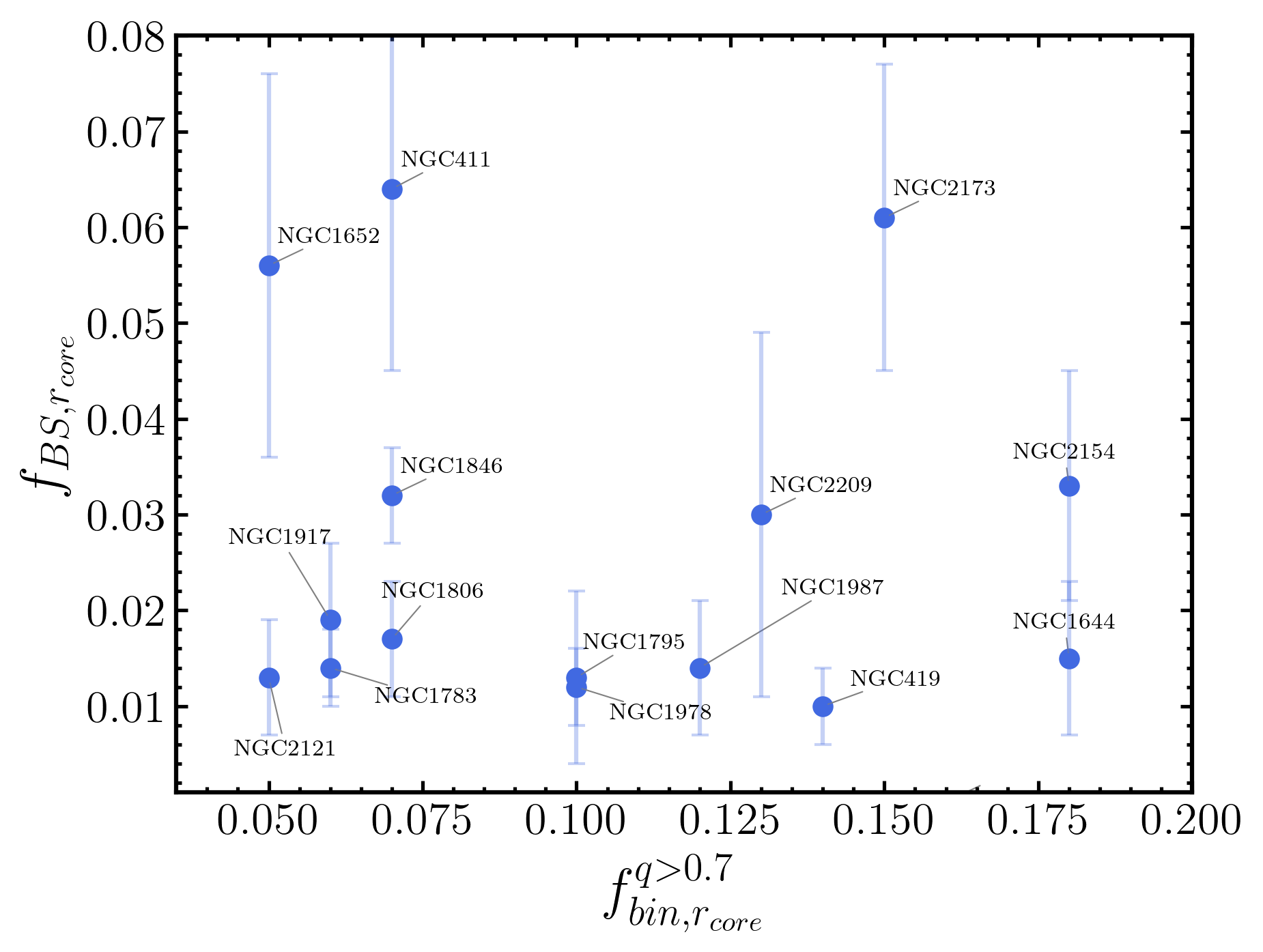} 
    \caption{ Fraction of candidate BSs as a function of the fraction of binaries with mass ratio, $q\geq0.7$ in the studied clusters.}
    \label{comparing_BSs}
\end{figure}

To investigate the BSs population, we followed the procedure described in \cite{cordoni2023} and illustrated in Fig. \ref{BSs} for NGC\,1806. The fraction of BSs was estimated using the formula:
\begin{equation}
    f_{\rm BS}=\frac{N^C-N^C_{field}}{N^D-N^D_{field}}
\end{equation}
where $N^C$ and $N^D$ represent the number of cluster member stars in regions C and D of the CMD in Fig. \ref{BSs}, respectively. $N^C_{field}$ and $N^D_{field}$ refer to the number of stars in the field region.  Region C is defined as the area enclosed between the zero-age MS isochrone and a fiducial line red-shifted by 0.2 color, capturing the BS candidates. For region D, we delimited the interval between the MS turn-off and one magnitude below the MS turn-off in a square region. 
The uncertainty on the fraction is determined in the same way as for the binary fraction, by employing Monte Carlo simulations.
All fractions are listed in Table \ref{tab2}.
The fraction of candidate BSs is shown as a function of the binary fraction in Fig. \ref{comparing_BSs}. We do not detect any significant correlation between these two parameters, consistent with the findings for Galactic OCs \citep{cordoni2023} and other MC clusters \citep{mohandasan2024}.
Moreover, there are no indications that clusters with differing densities tend to lie along distinct sequences as identified in \cite{cordoni2023}.
We calculated a Pearson correlation coefficient of -0.14 (p = 0.59) and a Spearman correlation coefficient of -0.12 (p = 0.66). These results indicate that there is no statistically significant correlation.

\section{Summary and Conclusions}

We have presented a comprehensive photometric study of binary populations in 16 star clusters from the MC using high-precision HST observations obtained with the UVIS/WFC3 and ACS/WFC cameras. Our analysis spans clusters with diverse ages, masses, and metallicities, providing valuable insight into the properties and distributions of binary systems in these environments.
\begin{itemize}

\item We derived the structural parameters of each cluster by fitting the density profiles with the EFF profiles. We provide for the community the core radii, $r_{50}$, and central densities by fitting the observed density profiles with the EFF profiles.

\item We measured the binary fractions over the entire field of view, within the core radius, and, when feasible, in the region between the core and $r_{50}$ radii. For binaries with mass ratios larger than 0.7, the inferred binary fractions span from 5\% in NGC\,2121 to 13\% in NGC\,2173, in line with values reported for other clusters in other MC clusters.
Using the methodology of \cite{muratore2026}, we estimate the binary fraction in two mass-ratio intervals (0.7–0.8 and 0.8–1.0), where field contamination is less severe.
We compute the ratio of the fractions in the two bins to serve as a proxy for the underlying mass-ratio distribution. The average of all ratios results in a flat distribution.
\item By combining our measurements with previous studies, including 14 MC clusters analyzed by \cite{mohandasan2024}, 78 Galactic OCs from \cite{cordoni2023}, and 67 Galactic GCs studied by \cite{milone2012, milone2016}, we have built a homogeneous data set of binary properties for 165 star clusters that span a wide range of ages, masses, and environments. This uniquely large sample enables us to explore how binary fractions depend on cluster properties.
We find no meaningful correlation between the binary fraction and the age of the cluster.  
We find a clear anti-correlation between the core binary fraction and total cluster mass, which supports previous studies on Galactic and Magellanic Cloud clusters. 

This behavior is consistent with the findings of \cite{ivanova2005}, who used Monte Carlo simulations to study binary-star populations in star clusters. They showed that GCs typically exhibit lower core binary fractions, resulting from the combined effects of stellar evolution and dynamical processes, including binary–single and binary–binary interactions.

\item We derived the MFs for each cluster, accounting for both completeness effects and the contribution of binary systems. We identified a correlation between the MF slopes and $r_{core}$, more generally between the positions of the clusters in log(age)–$r_{core}$ and the MF slopes. This behavior aligns with the expectations of \cite{elson1991}, who proposed that as clusters evolve and grow, their core radii adjust in response to the initial mass function. Specifically, clusters with a flatter IMF (i.e., containing relatively more massive stars) lose mass more quickly through stellar evolution, leading to faster and more substantial core expansion, whereas clusters with a steeper IMF maintain compact cores for longer periods.
Additionally, \cite{wilkinson2003} tested the LMC tidal field and a large primordial population of hard binaries to explain the expansion of the core. Their N-body simulations indicate that neither mechanism alone can account for the full observed spread. However, these authors evolved their simulations for only 1.7 Gyr; therefore, extending the evolution of a cluster to at least 2.5 Gyr (i.e. the maximum age in our sample) could provide additional insight into the role of primordial binaries and the IMF on the evolution of the core radius.

\item We identified and characterized candidate BSs along the MSs of our target clusters. Interestingly, we find no correlation between the binary fraction and the fraction of candidate BSs, a result that mirrors observations in Galactic OCs and some MC star clusters \citep{cordoni2023, mohandasan2024}. This lack of correlation suggests that while binary evolution likely contributes to BSs formation through mass-transfer and merger channels, other mechanisms, such as collisional formation or primordial rotation, may also play significant roles or that the relationship between binaries and BSs is more complex than simple proportionality.

\end{itemize}
\section*{Data availability}
\small All tables are available in electronic form on Research Data Unipd: \url{https://researchdata.cab.unipd.it/id/eprint/1861}.

\begin{acknowledgements}
We thank the anonymous referee for helpful comments that significantly improved the quality of the article.
This work has been funded by the European Union – NextGenerationEU RRF M4C2 1.1 (PRIN 2022 2022MMEB9W: "Understanding the formation of globular clusters with their multiple stellar generations", CUP C53D23001200006).
This research is based on observations made with the NASA/ESA Hubble Space Telescope obtained from the Space Telescope Science Institute, which is operated by the Association of Universities for Research in Astronomy, Inc., under NASA contract NAS 5–26555. These observations are associated with programs 12257, 14069, 10396, 9891, 10595, 15630, 12908.
\end{acknowledgements}

%
%
\bibliographystyle{aa}
\bibliography{mybib}

\end{document}